\documentclass[12pt]{amsart}

\usepackage{amsmath}
\usepackage{amssymb}
\usepackage{euscript}
\usepackage{varioref}

\def\cal{\mathcal}

\def\O{\Omega}
\def\be1{{\begin{equation}}}
\def\ee1{{\end{equation}}}
\def\Om{\Omega}\def\om{\omega}
\def\D{\Delta}

\def\RR{{\mathbb R}}
\def\CP{{\mathbb CP}}
\def\C{\mathbb C}

\def\a{\alpha}

\def\part{\partial}
\def\R{{\mathbb R}}

\def\cs{Chern-Simons }

\def\r{relativistic }
\def\ba{\begin{array}}
\def\ea{\end{array}}

\newtheorem{them}{Theorem}[section]
\newtheorem{lem}[them]{Lemma}

\newtheorem{coro}[them]{Corollary}

\newtheorem{pro}[them]{Proposition}

\numberwithin{equation}{section}

\begin{document}
\title[Solutions of a Toda system]
{Classification  of solutions of a Toda system in $\R^2$}
\author [J.Jost and G. Wang] 
{J\"urgen Jost and  Guofang Wang }
%\size{6}{10pt} \selectfont
 \address {Max-Planck-Institute
  for Mathematics in Sciences\\
 Inselstra\ss e 22-26\\
 D-04103 Leipzig}

\date{\today}
\begin{abstract} In this paper, we consider solutions of the following 
(open) Toda system (Toda lattice) for $SU(N+1)$
\[-\frac 12 \D u_i = \sum_{j=1}^N a_{ij} e^{u_j} \quad \text {in } \R^2,\] 
for $j=1,2,\cdots, N$, where $K=(a_{ij})_{N\times N}$ is the Cartan
matrix for $SU(N+1)$. We show that any solution $u=(u_1,u_2,\cdots, u_N)$
with \[\int_{\R^2}e^{u_i}< \infty, \quad i=1,2,\cdots, N,\]
can be obtained from a rational curve in $\C P^N$.

\end{abstract}\maketitle

\section{Introduction}
Let $N>0$ be an integer. The 2-dimensional (open) Toda system (Toda lattice)
for $SU(N+1)$
is the following system
\begin{equation}\label{1.1}
-\frac 12 \D u_i = \sum_{j=1}^N a_{ij} e^{u_j} \quad \text {in } \R^2,
\end{equation}
for $j=1,2,\cdots, N$, where $K=(a_{ij})_{N\times N}$ is the Cartan
matrix for $SU(N+1)$ given by
\[
\left(\begin{array}{rrrrrr}
2 & -1 & 0 & \cdots & \cdots & 0\cr
-1& 2 &-1& 0 & \cdot & 0\cr
0&-1&2&-1& \cdots & 0\cr
\cdots&\cdots&\cdots&\cdots&\cdots&\cdots\cr
0 &\cdots&\cdots& -1&2&-1\cr
0&\cdots&\cdots & 0 &-1&2\cr\end{array}\right).\nonumber
\]
(Here the factor $\frac 12$ comes from $\frac 12 \D u=u_{z\bar z}$.) 
System (\ref{1.1}) is a very natural generalization of the Liouville
equation
\begin{equation}\label{1.2}
-\D u =2e^{u},
\end{equation}
which is completely integrable, known from Liouville \cite{L}.
Roughly speaking, any solution of (\ref{1.2}) in a simply connected domain
arises  from a holomorphic function.
System (\ref{1.1}) is also completely integrable. All solutions
of (\ref{1.1}) in a simply connected domain
 arise from $N$ holomorphic functions,
see \cite{Ko, LS1,LS2, Dunnebook}. 
However, it is difficult to determine the precise form of these holomorphic
functions,  when we require additional (or boundary) conditions  for
(\ref{1.2}) or (\ref{1.1}).

Recently, Chen-Li \cite{ChenL1} classified all solutions of (\ref{1.2}) in $\R^2$
with 
\begin{equation}\label{1.3}
\int_{\R^2}e^{u}< \infty.
\end{equation}
Their result is very useful for  2-dimensional problems, especially
for the study of 
the Moser-Trudinger inequality and the mean field equation, see
\cite{djlw,  NT1}. To obtain their classification,
they used an energy inequality of Ding \cite{Ding} and the method of 
moving plane to show that
any solution has a rotational symmetry.
 Other proofs of the classification result were given by
Chou-Wan \cite{Chou-Wan} by using the complete integrability mentioned above
and complex analysis and by Chanillo-Kiessling \cite{CK} by using
an isoperimetric inequality and a global Pohozaev identity. The latter was 
applied to classify solutions of a class of Liouville type systems with
non-negative coefficients in \cite{CK}. 
See also the work of Chipot-Shafrir-Wolansky \cite{csw}. The methods used 
by Chen-Li and Chanillo-Kiessling  rely on the maximum principle.
Hence, it is difficult (or impossible) to apply their method
to study the similar problem for System (\ref{1.1}). 
We believe the method of Chou-Wan can be applied to study (\ref{1.1}).
In fact, we notice that a similar method was used by Bryant \cite{Bryant} in
his study of pseudo-metrics. We believe that his method can be adapted to
classify (\ref{1.1}) by using the Nevanlinna theory for  holomorphic curves
into $\CP^N$ instead of that for holomorphic functions. 
For the Nevanlinna theory for holomorphic curves into
$\CP^N$, see for example \cite{Fujimoto}.

System (\ref{1.1}) has a very close relationship with
a few geometric objects, holomorphic curves into
$\C P^N$, flat $SU(N+1)$ connections and harmonic sequences, see
for instance, \cite{Calabi, Lawson,GrH, GM, BW, EellsW, ChernW}.
To classify solutions of (\ref{1.1}), it is natural to seek 
the help of differential geometry.
Here, with such a help, we classify 
system (\ref{1.1}) with
\begin{equation}\label{1.4}
\int_{\R^2}e^{u_i}<\infty, \quad i=1,2,\cdots, N.
\end{equation}

\begin{them}\label{MainTheorem1} Any 
$C^2$ solution $u=(u_1,u_2, \cdots, u_N)$  of (\ref{1.1}) and (\ref{1.4})
has the following form
\begin{equation}\label{1.5}
u_i(z)=\sum_{j=1}^Na_{ij} \log\|\Lambda_j(f)\|^2
\end{equation}
for some rational curve in $\C P^N$. For the definition of
$\Lambda_k(f)$, see section 3 below.
\end{them}

Any rational curve in $\C P^N$ can be transformed to
\[
\phi_0(z)=[1,z, \cdots, \sqrt { \left(\begin{matrix} N \\ k \end{matrix} 
\right)}z^k, \cdots,
z^n],
\]
by a holomorphic isometry, which is an element of $PSL(N+1, \C)$. Hence 
the space of solutions of (\ref{1.1}) and (\ref{1.4}) is equivalent to
$PSL(N+1, \C)/PSU(N+1)$. The dimension of the solution space is $N^2+2N.$

Theorem \ref{MainTheorem1} can be restated in a geometric way as follows:

\begin{them}\label{MainTheorem2} 
Any totally unramified holomorphic map $\phi$ from $\C$ to $\CP^N$ 
satisfying the finite energy condition (\ref{energy}) below 
can be compactified to a rational curve.
\end{them}

Theorem \ref{MainTheorem2} is a generalization of the following 
well-known result:
Any totally unramified compact curve in $\C P^N$ is rational.

When $N=1$, Theorem 1.1 is just the classification result of Chen-Li.
%When $N=2$, system (\ref{1.1}) is
%\begin{equation}\label{1.6}\left\{\begin{array}{rcl}
%-\frac 12 \D u_1& =& \,\,2e^{u_1}-e^{u_2}\\
%-\frac 12\D u_2& =& -e^{u_1}+2e^{u_2}. \end{array}\right.
%\end{equation}
%Any solution of (\ref{1.6}) and (\ref{1.4}) has the following
%form:

%System (\ref{1.6}) satisfies all condition of Theorem in \cite{CK},
%except the non-negativity of coefficients. 
%Their result concludes that any
%solution of Liouville type system satisfying the conditions of 
%Theorem in \cite{CK}
%is of Liouville type, i.e., $u_1=u_2=$. 
%Hence the space of solutions is of dimension $3$.

The Toda system is of great interest not only in
geometry, but also in mathematical physics. One of our motivations to
study this system is the non-abelian Chern-Simons Higgs model,
in which non-topological solutions are solutions of a perturbed
Toda system. See \cite{Dunnebook, G, Yang, WZh, NT3}.

\section{Analytic aspects of the Toda system}
In this section, we  analyze
the asymptotic behavior of solutions of (\ref{1.1})-(\ref{1.4})
and obtain a global Rellich-Pohozaev identity. Since some results were presented
in our previous work \cite{JWa}, we only give an outline of ideas.
Similar methods were used in \cite{ChenL1, ChenL2, CK}. 

Let $I=\{1,2,\cdots, N\}$.
First, we have
\begin{lem} Let $u$ be a solution of (\ref{1.1})-(\ref{1.4}).
Then 
\[ u_i(z)=-\gamma_i \log|z|+a_i+O(|z|^{-1}) \quad \text{ for } |z| \text{ near }
\infty,\]
where
$a_i\in \R$ are some constants and $\gamma_i$ are given by
\[ \gamma_i=\frac{1}{\pi}\sum_{j=1}^Na_{ij}\int_{\R^2}e^{u_j}.\]
\end{lem}
\begin{proof}
First, one shows that
\begin{equation}\label{1}
 \max_{i\in I}\sup_{z\in \R^2} u_{i}(z)<\infty,\end{equation}
see \cite{JWa}. 
Set
\[v_i(z)=\frac 1 {\pi}\int_{\RR^2} (\log |x-y|-\log(|y|+1)) \sum_{j=1}^Ne^{u_j}
(y)dy,\]
for $i\in I$. (Note again that (\ref{1.1}) has a factor $\frac 12$.) 
 The potential analysis
implies
\begin{equation}\label{2}
  -\gamma_i\log|z|-C\le v_i(z) \le -\gamma_i\log|z|+C,
  \end{equation}
for  some constant $C> 0$, see \cite{ChenL2}.
Clearly $u_i-v_i$ is a harmonic function. Hence
(\ref{1}) and (\ref{2}) imply that $u_i-v_i=c_i$ for some constant $c_i$. 
That is, $u$
has the following representation formula
\[ u_i(z) =\frac 1 {\pi}\int_{\RR^2} (\log |x-y|-\log(|y|+1)) \sum_{j=1}^Ne^{u_j}
(y)dy +c_i.\]
The above results  and (\ref{1.4}) imply  
\begin{equation}\label{3}
\gamma_i > 2, \quad i\in I.\end{equation}
Furthermore, we can show that
\[u_i=\gamma_i\log|z|+a_i+O(|z|^{-1}).\]
See, for example, \cite{WZ}.
\end{proof}

Next, we have a global Rellich-Pohozaev identity for our system (\ref{1.1}).
Such an  identity was obtained in \cite{CK} for a Liouville
type system with nonnegative entries. 
Similar arguments
work for our case, see \cite{JWa}. Here we give another proof
that is similar to the spirit of proof of the Main Theorems.

\begin{pro} Let $u$ be a solution of (\ref{1.1}) and (\ref{1.4}).
Then we have
\begin{equation}
\label{4}
\sum_{j,k=1}^N a^{jk}(4\gamma_k-\gamma_j \gamma_k)=0,\end{equation}
where the matrix $(a^{ij})$ is the inverse of the Cartan matrix $(a_{ij})$.
\end{pro}
\begin{proof}
Set
\[f=\sum_{j,k=1}^N a^{jk}\{(u_{k})_{zz}-\frac 12 (u_j)_z\cdot (u_k)_z\}.\]
We can check that $f$ is a  holomorphic function as follows:
\[\begin{array}{rcl}
f_{\bar z} &=& 
\frac 12\sum_{j,k=1}^N a^{jk}\{(\D u_{k})_{z}- (u_j)_z\cdot \D u_k\} \\
&=& -\sum_{j,k,l=1}^N a^{jk}\{ a_{kl} e^{u_l}(u_l)_z-
a_{kl}(u_j)_z e^{u_l}\}\\
&=&-\sum_{j=1}^N(e^{u_j}(u_j)_z-e^{u_j}(u_j)_z)\\
&=& 0.\end{array}\]
In the first equality we have used the symmetry of the matrix $(a^{ij})$.
Using Lemma 2.1, we have the following expansion of $f$ near infinity
\[ \frac 18  \frac 1{z^2} \sum_{j,k-1}^N a^{jk}(4\gamma_k-\gamma_j \gamma_k)
+\frac {c_{-3}} {z^3}+ \cdots.\]
Hence, $f $ is a constant (zero, in fact) and
\[
 \sum_{j,k=1}^N a^{jk}(4\gamma_k-\gamma_j \gamma_k)=0.\]
\end{proof}

\section{Geometric aspects of the Toda system}

In this section, we recall some relations between
the Toda system and various geometric objects,
flat connections, holomorphic curves into $\C P^N$
and harmonic sequences.
Furthermore, we relate the mild singularities of solutions of the
Toda system with the holonomy of the corresponding flat connections.

\subsection{From solutions of Toda systems to flat connections}
Let $\Omega$ be a simply connected domain and 
$u=(u_1,u_2, \cdots, u_N)$  a solution of (\ref{1.1}) on $\Omega$. Define
$w_0, w_1,w_2,\cdots, w_N$ by the following relations
\begin{equation}
\label{2.1} u_i=2w_i-2w_0 \quad\text{ for } i\in I
\text { and }
\sum_{i=0}^Nw_i=0.\end{equation}
It is easy to check that
$w_0, w_1,\cdots, w_N$ satisfies
\begin{equation}\label{2.3}\left\{ \begin{array}{lll}
- \D w_0=2(w_0)_{z\bar z}& =& e^{w_1-w_0}\\
-\D w_1 =2(w_1)_{z\bar z}& =& - e^{w_1-w_0}+ e^{w_2-w_1}\\
 \cdots  & \cdots&\cdots\\
-\D w_N =2(w_N)_{z\bar z} &=& -e^{w_N-w_{N-1}}.\\
\end{array} \right.\end{equation}

It is well-known that (\ref{2.3}) is equivalent to
an integrability condition 
\begin{equation}\label{integ}
{\cal U}_{\bar z}-{\cal V}_{ z}= [{\cal U}, {\cal V}]
\end{equation}
of the
following two equations
\begin{equation}\label{2.4}
\phi^{-1}\cdot\phi_z= {\cal U}\end{equation}
and 
\begin{equation}\label{2.5}
\phi^{-1}\cdot\phi_{\bar z} ={\cal V,}\end{equation}
where
\[{\cal U}=
\left(\begin{matrix} (w_0)_{z} & &&\\
& (w_1)_z && \\
&& \cdots & \\
&&& (w_N)_z \\
\end{matrix}\right)+
\left(\begin{matrix} 0 & e^{w_1-w_0} &&\\
 & 0  && \\
 && \cdots &e^{w_N-w_{N-1}} \\
&&& 0 \\
\end{matrix}\right)
\]
and
\[{\cal V} = 
-\left(\begin{matrix} (w_0)_{\bar z} & &&\\
& (w_1)_{\bar z} && \\
&& \cdots & \\
&&& (w_N)_{\bar z} \\
\end{matrix}\right)-\left(\begin{matrix} 0 & &&\\
e^{w_1-w_0} & 0  && \\
 && \cdots & \\
&&e^{w_N-w_{N-1}}& 0 \\
\end{matrix}\right)
\]

 Hence, from a solution of (\ref{1.1}) (or equivalently (\ref{2.3}))
we first get a one-form $\a={\cal U}dz+{\cal V}d\bar z$.  Then,
with the help of  the Frobenius Theorem,  we 
obtain a map $\phi:\O\to SU(N+1)$ such that
\[\a=\phi^{-1}\cdot d\phi.\]

It is clear that $\a$ (or $d+\a$) is a flat $SU(N+1)$ connection on the trivial
bundle $\Omega\times \C^{N+1}\to \Omega,$ i.e.,
$\a$ satisfies the Maurer-Cartan equation
\[ d\a+\frac 12[\a,\a]=0\]
which is  equivalent to the integrability condition
(\ref{integ}), hence (3.2).
\begin{lem}\label{lem2.1}
$\phi$ is determined upto an element of $SU(N+1)$.
That is, any two 
$\phi_1, \phi_2:\Om \to SU(N+1)$ with $\phi_1^{-1}d\phi_1=
\phi_2^{-1}d\phi_2=\alpha$ satisfy
\[ \phi_1=g\cdot \phi_2,\]
for some element $g\in SU(N+1)$.
\end{lem}

We call $\phi$ and $\a$ a {\it Toda map} and {\it Toda form} respectively.

\subsection{Holonomy of a singular connection}
Now consider an $SU(N+1)$ connection $\a$ on the punctured disk $D^*$.
We can define its holonomy as in \cite{SS}.
When $\Omega=D^*$ is not simply connected, we cannot apply
the Frobenius theorem directly and have to 
consider  the {\it holonomy}. Let $(r,\theta)$ be the polar
coordinates. Write $\a$ as $\a=\a_rdr+\a_{\theta} d\theta$.
$\a_r$ and $\a_\theta$ are $su(N+1)$-valued. For any given $r\in (0,1)$, the 
following initial value problem,
\[ \frac {d\phi_r}{d\theta}+\a_{\theta}\phi_r =0, \quad \phi_r(0)=Id,\]
has a unique solution $\phi_r(\theta)\in  SU(N+1)$. Here $Id$ is the identity matrix.

\begin{lem} If $\a$ is a flat connection on $D^*$,
then $\phi_r(2\pi)$ is independent of $r$.
\end{lem}

Let $h_\a$ denote $\phi_r(2\pi)$. $h_\a$ is called the {\it holonomy}
of $\a$.

\noindent{ \it Remark.} Here, we use a slightly different
 definition of holonomy. The usual holonomy is defined by
 the conjugacy class of $h_\a$, which is invariant
 under gauge transformations.
 
\begin{pro}
Let $u=(u_1,u_2,\cdots,u_N)$ be a solution of (2.1) with
\[u_i(z)=-\mu_i\log |z| +O(1), \quad \text{  near } 0.\]
If $\mu_i<2$ for $i\in I$, 
then the corresponding flat connection $\a$ has holonomy
\[
h_\a= \left(\begin{matrix}e^{2\pi i \beta_0} & &&\\
&  e^{2\pi i\beta_1} && \\
&& \cdots & \\
&&& e^{2\pi i \beta_N }\\
\end{matrix}\right),\]
where $\beta_0,\beta_1,\cdots,\beta_N$ are determined by
\begin{equation}\label{pro3.1.1} \beta_i-\beta_0=\frac 12 \mu_i \quad  (i\in I)\quad
\text { and }
\quad \sum_{j=0}^N\beta_j=0.\end{equation}
\end{pro}
\begin{proof} Define $w_i$ by (\ref{2.1}). From the assumption, we
have
\[ w_i=-\beta_i\log|z|+O(1), \quad \text {near } 0.\]
A direct computation shows that
\[{\mathcal U} =
 \frac 1{2z}\left(\begin{matrix}  -\beta_0 & &&\\
&  -\beta_1 && \\
&& \cdots & \\
&&& -\beta_N \\
\end{matrix}
 \right)+o(\frac1{|z|}),\]
 where $o(\frac1{|z|})$ means that a matrix $(b_{ij})$ with
 entries  satisfying $|z|b_{ij}\to 0$ as $|z|\to 0$.
 Here, we have used the condition that $\mu_i<2$ for any $i\in I$.
 Similarly,
 \[{\mathcal V} =
 \frac 1{2\bar z}\left(\begin{matrix}  \beta_0 & &&\\
&  \beta_1 && \\
&& \cdots & \\
&&& \beta_N \\
\end{matrix}
 \right)+o(\frac{1}{|z|}).\]
 Hence,
\[\a_\theta=\sqrt{-1}\left(\begin{matrix}  \beta_0 & &&\\
&  \beta_1 && \\
&& \cdots & \\
&&& \beta_N \\
\end{matrix}
 \right)+o(1).\]
Now it is easy to compute the holonomy of $\a$. 
\end{proof}

\subsection{From solutions of (\ref{1.1}) to  holomorphic curves}
When we have a Toda map $\phi:\Omega \to SU(N+1)$ from a solution of 
the Toda system, 
we can get a
harmonic sequence  as follows.
First, define $N+1$ $\C^{N+1}$-valued functions $\hat f_0, \hat f_1,\cdots
\hat f_N$ by
\[(\hat f_0, \hat f_1,\cdots
\hat f_N)=\phi\cdot
\left(\begin{matrix} e^{w_0} & & &\\
& e^{w_1} & &\\
& & \cdots &\\
& & & e^{w_N}\\ \end{matrix}\right).\]
Let $f_i$  denote the map into $\CP^N$ obtained from $\hat f_i$.
It is easy to check that
 $f_i$ is a harmonic map and satisfies
\begin{equation}
\begin{array}{lll}
(\hat f_k)_z& =& \hat f +a_k\hat f_k,\\ 
(\hat f_k)_{\bar z} & =& b_k\hat f_{k-1},\\
\end{array}\label{2.3.1}
\end{equation}
where \[a_k=(\log |\hat f_k|^2)_z=(e^{2w_k})_z \text{ and }
b_{k-1}=-|\hat f_k|^2/|\hat f_{k-1}|^2=-w^{2(w_k-w_{k-1})}.\]
Here we assume that  $\hat f_{-1}=\hat  f_{N+2}=0$.
Hence, $f_0$ is a holomorphic map and $f_{N+1}$ is an
anti-holomorphic map into $\CP^N$.
In fact, (\ref{2.3.1}) is the Frenet frame of the holomorphic
map $f_0$, see \cite{GrH} or below. 
Furthermore, $f_0$ is unramified in $\Om.$
For the definition of the ramification index, see \cite{GrH} or below.

\subsection{From a curve to a solution of the Toda system}
 From a nondegenerate (i.e. not contained in  a proper projective subspace
 of $\C P^N$) holomorphic curve $f_0$ into $\CP^N$, one can get 
a family of associated curves into various Grassmannians as follows.
Lift $f_0$ locally to $\C^{N+1}$ and denote the
lift by $v=(v_0,v_1,\cdots, v_N)$.
Hence, $f_0=[v_0,v_1,\cdots, v_N]$. 
The 
 {\it $k$-th associated curve} of $f_0$ is defined by
 \[\begin{array} {lll}f_k:\Om &\to&  G(k+1, n+1)\subset \CP^{N_k}\\
 f_k(z)& =& [\Lambda_k],
 \end{array}\]
 where
 \[\Lambda_k=v(z) \wedge v'(z)\wedge \cdots \wedge v^{(k)}(z).\]
See for example \cite{GrH}. Here $N_k=\left( \begin{matrix} {N+1} \\ k+1
\end{matrix}\right).$

Let $\om_k$ be the Fubini-Study metric on $\C P^{N_k}$. 
The well-known (infinitesimal) Pl\"ucker formula is
\begin{equation}
\label{infin}
 f_k^*(\omega_k)=\frac {\sqrt{-1}}2
\frac{\|\Lambda_{k-1}\|^2\cdot \|\Lambda_{k+1}\|^2}
{\|\Lambda_{k}\|^4}dz\wedge \bar z,\end{equation}
which implies
\begin{equation}
\label{plueker}
 \frac {\partial ^2}{\partial z\partial \bar z}
\log\|\Lambda_k\|^2=
\frac{\|\Lambda_{k-1}\|^2\cdot \|\Lambda_{k+1}\|^2}
{\|\Lambda_{k}\|^4}, \quad\text{ for } k=1,\cdots, N,
\end{equation}
where $\|\Lambda_0\|^2=1$ and $\|\Lambda_N\|=\det (f,f',\cdot, f^{(N)})$.
By choosing the normalization $\|\Lambda_N\|=1$
(we can do this when we lift $f$), we can identify
(\ref{plueker}) with the Toda system (\ref{1.1}) as follows.
By setting
\[v_k=\log\|\Lambda_k\|^2,\]
 system (\ref{plueker}) becomes
 \begin{equation}\label{toda2}
 -\frac12 \D v_i= \exp \{\sum_{j=1}^N a_{ij} v_j\}.\end{equation}
Clearly, (\ref{toda2}) is equivalent to (\ref{1.1})
by setting
\[u_i=\sum_{j=1}^N a_{ij} v_j.\]
%The relation between the Pl\"ucker formula for holomorhic curves
%with the Toda system was observed in \cite{GM}, but at least went back
%to \cite{Calabi, Lawson}, see also.

For any curve $f:\Om\to \CP^n$, the {\it ramification index}
$\beta(z_0)$ at $z_0\in \Om$ is defined by the unique real number 
such that 
\[f^*\omega=\frac{\sqrt{-1}}2|z-z_0|^{2\beta(z_0)}\cdot h(z)\cdot dz\wedge
d\bar z\]
with $h$ $C^\infty$ and non-zero at $z_0$, where
$\omega$ is the K\"ahler form of the Fubini-Study metric on $\CP^n$.
For other definitions, see \cite{GrH}. 
$f$ is  unramified if for any  $z_0\in \Omega$ 
the ramification index $\beta(z_0)$ vanishes. Hence,
a solution of the Toda system (\ref{1.1}) corresponds to
an unramified holomorphic curve.

Let $f:\C\to \C P^N$ be a holomorphic curve and $f_k$ its
$k$-th associated curves. The finite energy condition is defined
by
\begin{equation}\label{energy}
\int_{\R^2}f^*(\om_k) < \infty, \quad \text{ for any k }\in I.
\end{equation}
(\ref{energy}) means that the area of the $k$-th associated curve is bounded.

\section{Proof of Main Theorems}
Now we start to prove our main Theorems.

\noindent{\it Proof of Theorem 1.1.}
Let
\[ v_i(z)= u_i(\frac{\bar z}{|z|^2})-4\log|z|, \quad i\in I.\]
$v=(v_1,v_2,\cdots,v_N)$ satisfies (\ref{1.1}) on
$\R^2/\{0\}$. 
Applying Lemma 2.1,
 we have
\[v_i(z)= (\gamma_i-4) \log|z| +O(1) \quad \text{ near } 0.\]
Hence, using (2.3) Proposition 3.3 implies that the holonomy of the
corresponding 
Toda form of $v$ is
 \[
h_\a=\left(\begin{matrix} e^{2\pi i \beta_0} & &&\\
&e^{2\pi i \beta_1} && \\
&& \cdots & \\
&&& e^{ 2\pi i \beta_N }\\
\end{matrix}\right),\]
where $\beta_0,\beta_1,\cdots,\beta_N$ are determined by
\[ \beta_i-\beta_0=\frac12(\gamma_i-4 )\quad \text { and }
\quad \sum_{j=0}^N\beta_j=0.\]

Now we know that the holonomy is trivial, i.e.,
$h_\a$ is  the identity matrix, which clearly implies
\[\beta_i = 2 \quad \text{mod } {\mathbb Z} \quad \text{ for }i=0,1,
\cdots, N.\]
Hence, we have 
\[\gamma_i = 2 \quad \text{mod } {\mathbb Z} \quad \text{ for any }i\in I,\]
which, together with (2.3), implies that $\gamma_i\ge 4$ for any
$i\in I$. Thus,
$4\gamma_k-\gamma_j\gamma_k\le 0$ for any $j, k\in I$. On the other hand,
one can check that  the matrix $(a^{ij})$ does not admit negative entries.
In fact,
a direct computation shows that
\[a^{ij}=\frac{i(N+2-j)}{N+2}, \quad \text { for } i,j \le \{\frac{N+2}{2}\},\]
where $\{b\}$ means the least integer larger than or equal to $b$.
Other entries are determined by an obvious symmetry.
Altogether, we obtain
\[\sum_{j, k=1}^N a^{jk}(4\gamma_k-\gamma_j\gamma_k)\le 0,\]
and the equality holds if and only if $\gamma_i=4$ for any $i\in I$.
Applying the global Rellich-Pohozaev identity (2.4), we have
\[\gamma_i=4 \quad \text{ for any } i\in I.\]

Hence, $v_i$ is bounded near $0$. The elliptic theory implies
that $v_i$ is smooth. From the  discussions presented in 
Section 3, it follows that the corresponding
holomorphic curve $f$ can be viewed as an 
unramified map from ${\mathbb S}^2$ to $\C P^N$,
 hence this curve is a rational curve, namely
\[
f=[1,z, \cdots, \sqrt { \left(\begin{matrix} N \\ k \end{matrix} 
\right)}z^k, \cdots,
z^N],
\]
up to a holomorphic isometry, an element in $PSL(N+1,\C)$.
Now we can investigate any solution of (\ref{1.1}) and (\ref{1.4})
as in subsection 3.4.
 From a holomorphic curve $f$, we get the $k$-th associated curves 
$f_k$ and $\Lambda_k$. The solution of (1.1) $u=(u_1, u_2,\cdots, u_N)$ is
given by
\[ u_i=\sum_{j=1}^N a_{ij}\log \|\Lambda_j\|^2.\]
This completes the proof.
%{\hfill$\Box$}

\medskip

\noindent{\it Proof of Theorem 1.2.} From such a holomorphic
curve $f: \C \to \C P^N$, we get a solution $u$ of (1.1). It is clear 
that the condition (\ref{energy}) implies that $u$ satisfies
(\ref{1.4}). As in the proof of Theorem 1.1,  $f$ can be extended
to a curve from ${\mathbb S}^2$ to $\C P^N,$ which
is totally unramified. Hence, it is a rational curve.
%{\hfill$\Box$}

\begin{coro} The space of solutions of (\ref{1.1}) and (\ref{1.4}) is
$PSL(N+1,\C)/PSU(N+1)$.
\end{coro}

\begin{proof} It follows from Theorem 1.1 and Lemma 3.1.
\end{proof}

\begin{coro} Any solution $u=(u_1,u_2,\cdots, u_N)$ of (\ref{1.1})
and (\ref{1.4}) satisfies
\[ \frac 1{\pi}\sum_{j=1}^N a_{ij} \int_{\R^2} e^{u_j}=4.\]
In particular, if $N=2$, then
\[2\int_{\R^2}e^{u_1}=2\int_{\R^2}e^{u_2}=8\pi.\]
\end{coro}

\end{document}